# *PAUNet: Precipitation Attention-based U-Net for rain prediction from satellite radiance data*

**P. Jyoteeshkumar Reddy***

Commonwealth Scientific and Industrial Research Organisation Environment Hobart, TAS, Australia.
jyoteesh.papari@csiro.au

**Harish Baki***

Geosciences and Remote Sensing Civil Engineering and Geosciences Delft University of Technology, Delft, The Netherlands.
h.baki@tudelft.nl

**Sandeep Chinta**

Center for Global Change Science Massachusetts Institute of Technology Cambridge, Massachusetts, USA.
sandeepc@mit.edu

**Richard Matear**

Commonwealth Scientific and Industrial Research Organisation Environment Hobart, TAS, Australia.

**John Taylor**

Commonwealth Scientific and Industrial Research Organisation Data61 Canberra, ACT, Australia.

*\* Equal contribution by the authors*

**Abstract**

This paper introduces Precipitation Attention-based U-Net (PAUNet), a deep learning architecture for predicting precipitation from satellite radiance data, addressing the challenges of the Weather4cast 2023 competition. PAUNet is a variant of U-Net and Res-Net, designed to effectively capture the large-scale contextual information of multi-band satellite images in visible, water vapor, and infrared bands through encoder convolutional layers with center cropping and attention mechanisms. We built upon the Focal Precipitation Loss including an exponential component (e-FPL), which further enhanced the importance across different precipitation categories, particularly medium and heavy rain. Trained on a substantial dataset from various European regions, PAUNet demonstrates notable accuracy with a higher Critical Success Index (CSI) score than the baseline model in predicting rainfall over multiple time slots. PAUNet's architecture and training methodology showcase improvements in precipitation forecasting, crucial for sectors like emergency services and retail and supply chain management.

**1. Introduction**

The accurate prediction of weather patterns, particularly extreme weather events, is critical in today's era of climate change, where unpredictable and severe meteorological occurrences are becoming more frequent [1]. Specifically, the ability to forecast precipitation accurately is of paramount importance, as it directly impacts numerous sectors such as agriculture, transportation, and emergency management. Precipitation events, ranging from light rain to severe storms, have profound implications for water management, transport, and public safety. One critical challenge with precipitation is to forecast the spatio-temporal variations accurately and rapidly [2, 3].

The weather prediction community has greatly benefited from the recent advances in deep learning. Multiple organizations, such as Google [4], NVIDIA [5], Microsoft [6], and others, have initiated the development of AI-based weather models. Specifically, Google developed convolutional neural network-based architectures for precipitation prediction, such as MetNet-1 [7], MetNet-2 [8] and MetNet-3 [9]. Several researchers also worked on this problem of precipitation forecasting using different architectures across various regions and with multiple datasets [10–14].

The Weather4cast 2023 competition [15–17] had three different challenges: Core, Nowcasting, and Transfer Learning. This paper presents our approach to this competition, which mainly included the incorporation of extracting the large-scale contextual information in the encoding part of our network, and we updated the FPL (Focal Precipitation Loss) to e-FPL (exponential - FPL) loss to efficiently focus across various thresholds (light-moderate-heavy) of precipitation, which are described in detail in the later sections. The data used and the methods implemented are detailed in sections 2 and 3, respectively. The results are presented in section 4. Section 5 provides a brief discussion and conclusion.

## 2. Data

The Weather4cast 2023 competition [15–17] presented an ambitious challenge in weather forecasting using satellite data, now enriched with various leaderboards to test different predictive models. The Core (Nowcasting) task is to predict the exact amount of rainfall over the next eight (four) hours, divided into 32 (16) time slots, based on a sequence of four-time slots from the preceding hour. This sequence comprised four 11-band spectral satellite images, each covering a 15-minute interval. These images spanned visible (VIS), water vapor (WV), and infrared (IR) bands, mapping a spatial area of approximately 12km x 12km per pixel. A unique aspect of this challenge was transforming these satellite images into high-resolution rain rate predictions based on ground-radar reflectivity, with each output image pixel corresponding to a finer spatial area of about 2km x 2km. This task required forecasting future precipitation and a super-resolution component due to the lower spatial resolution of the satellite data compared to the ground-radar output.

For the Core Challenge, the task was to forecast weather 8 hours into the future for 'known regions' (boxi_0015, boxi_0034, boxi_0076, roxi_0004, roxi_0005, roxi_0006, roxi_0007), covering the years 2019 and 2020. The dataset for this challenge was drawn from 10 European regions with distinct precipitation characteristics, with seven regions providing extensive training data for these two years. In contrast, the Nowcasting leaderboard introduced a different dimension by requiring a prediction of weather conditions 4 hours into the future for the same regions and years, aiming to explore the accuracy of shorter-term predictions. The Transfer Learning Challenge added an extra layer of complexity. It included satellite data for three additional regions (roxi_0008, roxi_0009, and roxi_0010) for 2019 and 2020 and extended to all ten regions for 2021, presenting both spatial and temporal transfer learning challenges. The datasets for the competition included both input satellite radiances and output ground-radar rain rates presented in 252x252 pixel patches. However, the satellite images had a spatial resolution approximately six times lower than the ground radar data, adding a compelling super-resolution aspect to the competition. This disparity meant that a 252x252 pixel ground radar patch corresponded to a central 42x42 pixel region in the satellite images, with the surrounding area providing ample contextual information for accurate weather prediction. This aspect of the Weather4cast 2023 competition to include higher resolution output posed an advanced challenge, requiring the effective downscaling and interpretation of coarser satellite data to match the finer resolution of ground-radar measurements. A total of 228,928 samples across seven regions were used for training, and the validation set contained 840 samples. The average critical success index (CSI) across five rain thresholds (0.2, 1, 5, 10, and 15) is considered as an evaluation metric. The architecture used in our approach is detailed in the next section.

## 3. Methods

### 3.1. Model Architecture

PAUNet (Precipitation Attention based U-Net) is developed based on the U-Net [18], Res-Net [19] architectures and attention mechanism [20] by incorporating appropriate modifications as necessary for the present study's problem, as shown in figure 1(a). The network is mainly built using 2D convolutions, 2D transposed convolutions/deconvolutions, and multi-head attention operations. All the convolution and deconvolution layers are non-linearly activated with ReLU activation. Firstly, the four-time samples of the input satellite multi-band images are concatenated across the channel dimension. The concatenated input is passed through the convolution block. The convolution block is designed based on the Res-Net, which consists of two 2D convolutions with 64 filters of size three operating on the residuals of the spatial features across the channels. The extracted residual spatial features are then added with input features obtained from the 2D convolution operation with 64 filters of size 1. Following the previous studies' approach of focusing on the center of the domain [15, 21, 22], we then applied the center crop to extract the critical spatial features associated with the target domain, i.e., 126×126 central part in the 252×252 contextual spatial features. Further, the convolution block and center crop are applied in the next step. Now, the spatial features represent the 4× coarsened target domain dimension (63×63) with large-scale contextual information across multiple channels. These large-scale contextual spatial features are concatenated with the 2× center cropped (63×63) input large-scale satellite multi-band images (252×252).

The encoded output is then passed to the attention layer with a single head of size 64. The attention layer enables the model to focus on the different regions of the encoded output, better capturing the relationship between extracted spatial features [20]. The attention output is then downscaled using the 2D deconvolution layer with 128 filters of size 3. The deconvolution layer performs the downscaling with learnable weights rather than fixed weights in interpolation-based convolution operations [11]. The downscaled data is passed to the convolution block to extract high-resolution spatial features independently across multiple channels. Next, the deconvolution and convolution block operations are repeated to extract the high-resolution target spatial features of precipitation for 32- (or) 16-time steps ahead. In the final layer, a 2D convolution operation with 32 (or) 16 filters of size three non-linearly maps the previous step output to the target domain precipitation data for 32- (or) 16-time steps ahead. The L1 regularization with 1e-5 regularization factor is considered for all the convolution and deconvolution layers in the network to avoid overfitting. The PAUNet consist of around 2.3M parameters in total.

### 3.2. Training procedure

In model training, the first step is data preprocessing. All the input multi-band satellite and radar precipitation data is converted to Zarr format chunking along the sample/batch dimension (i.e., 1×252×252×channels (44 for input and 32/16 for output)). Zarr chunking allows the selected sample to be loaded rapidly using CPUs in parallel as a part of a Dask cluster (Fig. 1(b)). The input data is normalized (min-max) over each multi-band satellite channel train sample, respectively. The radar precipitation data distribution is highly skewed with many no-rain samples. To overcome this, following previous studies, we have only considered the samples with at least rain in 100 grid points over the selected time sequence (16 or 32) [22]. Out of the overall 228,928 samples, 69,693 samples without rain were removed during the training stage. Further, the precipitation data is log-transformed and normalized with maximum (here considered 128 as per the previous studies), i.e., $\log_{10}(1+x)/\log_{10}(1+128)$, and then clipped all the values to the range 0-1. All values below 0.2 are considered to be zero due to the radar data artifacts, as suggested by the previous studies [15]. Even then, most values lie in the range 0.2-2. Hence, to focus the training across all categories of precipitation (0.2-1,1-5, 5-10, 10-15, and >15), we used the Focal Precipitation loss [23] but with the exponential weighting based on the precipitation rate, which we define as exponential Focal Precip Loss (e − FPL).

$$e - FPL = [\min(\beta e^{\alpha x}, 128)] \times [MAE(x', y')] \times [(1.01 - e^{-MAE})^\gamma] \quad (1)$$

Where x is the target precipitation data, $x'$ is the log transform normalized target precipitation data, and $y'$ is the model predicted output. We have performed several experiments with e-FPL and found that α in the range 0.5-0.6 with $\beta = 0.1$ and $\gamma = 0.5$, performs well in forecasting the rain across all the considered categories. In this study, we considered the α = 0.525, trained the PAUNet until the validation loss converged, and then restarted the training with α = 0.5 to emphasize training towards moderate to heavy precipitation categories. The models are trained using Adam optimizer with an initial learning rate of $10^{-3}$ until ten epochs, and then exponential decay of the learning rate is followed by a decay rate of $2.5\times10^{-4}$ to a minimum of $10^{-5}$. The restart runs are started again with the learning rate $10^{-3}$ for starting five epochs and then exponential decay to a minimum of $10^{-5}$. The model's loss curves show consistent training with the epochs and the convergence at the end, which shows no evidence of overfitting (Fig. 2). The mismatch between loss curves between the first run and the restart run is due to the change in α from 0.525 to 0.5.

a)

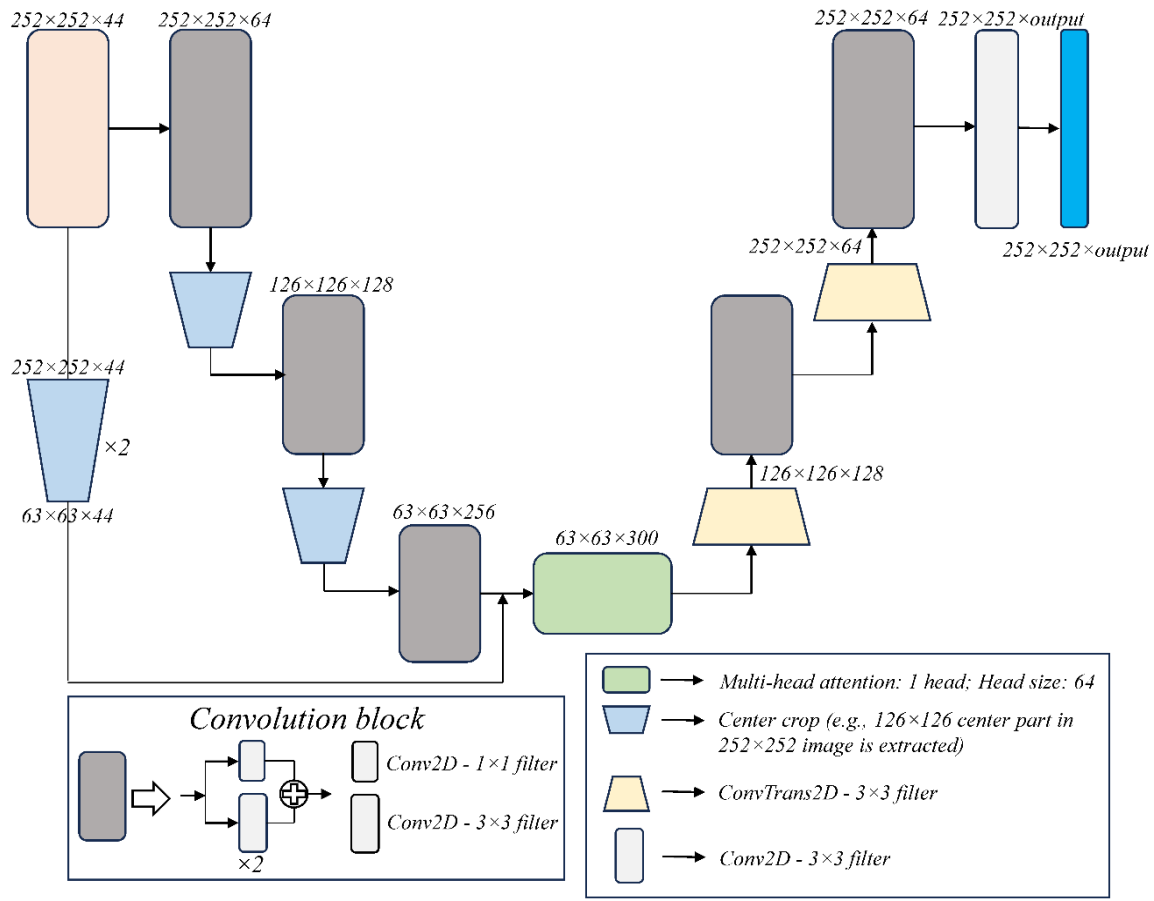

b)

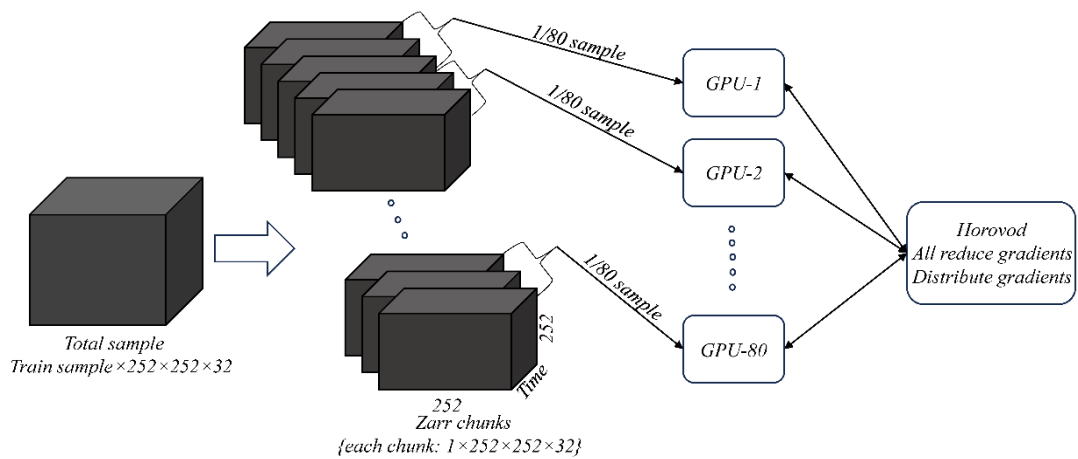

Fig. 1. a) An overview of PAUNet architecture. b) An overview of CPU/GPU parallelization of the training procedure. Parallel CPU data loading using Zarr chunking and Dask and parallel GPU distributed training using Horovod.

The models are developed using the TensorFlow package [24] and trained using a data-distributed parallel training framework using the Horovod package [25] (Fig. 1(b)). The distributed model training is performed across 80 Nvidia V100 GPUs, each with 32 GB memory, on the Gadi supercomputer at the National Computational Infrastructure in Canberra.

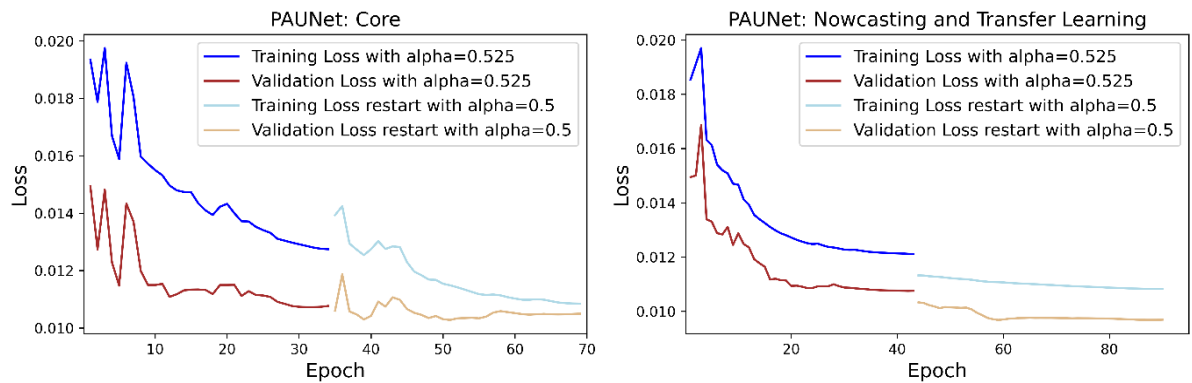

Fig. 2. Loss curve of PAUNet model trained for core (32 time-step forecasting, i.e., 8h) (left) and Nowcasting and Transfer learning (16 time-step forecasting, i.e., 4h) (right) challenge, respectively. The mismatch between loss curves between the first run and the restart run is due to the change in α value in the e-FPL from 0.525 to 0.5.

## 4. Results

Two PAUNet models were separately trained, one with 32 outputs for the core challenge and another with 16 outputs for nowcasting and transfer learning challenges. The learning curves of the trained models are presented in Figure 2, which clearly shows small variations in the validation loss across different epochs. Based on this ambiguity in selecting the best model, several models saved at different training epochs are used to obtain the predictions on the test dataset and are uploaded to the leaderboard. From the test scores, the first model trained for 53 epochs is obtained as the best model for the core challenge, while the second model trained for 62 and 90 epochs is found to be the best for the nowcasting and transfer learning challenges, respectively.

Table 1 presents the evaluation metrics of the best models obtained in addressing the core, nowcasting, and transfer learning challenges on the validation dataset. The CSI reflects the models' ability to accurately predict events at thresholds 0.2, 1, 5, 10, and 15, providing insights into the models' performance under different precipitation categories. The overall CSI offers a consolidated view of the predictive accuracy across all thresholds. Additionally, the Heidke Skill Score (HSS) is included, providing an assessment of the models' skill in relation to random chance. From these scores, it is evident that the overall CSI is dominated by the contribution from the 0.2 threshold. The prediction scores decrease as the threshold increases, with the CSI at 15 mm/hr threshold being the least contributor. This signifies the necessity of improving the model's performance at higher thresholds to increase the overall precipitation prediction capability. Another observation is that the evaluation scores are higher in nowcasting and transfer learning compared to the core challenge, because the nowcasting and transfer learning challenges predict precipitation for only 4 hours (16 time-steps), while the core challenge forecasts precipitation for 8 hours (32 time-steps).

Table 2 presents the performance evaluation scores of the best models on the test dataset and compares them against the official leaderboard baseline scores. Notably, the PAUNet demonstrates substantial improvement over the baseline scores across all three challenges. In core and nowcasting, the PAUNet outperforms the baseline by 10% and 18%, respectively. In the transfer learning challenge, the PAUNet exhibits 25% enhancement compared to the official baseline. The results underscore the efficacy of the PAUNet model in advancing predictive accuracy, highlighting its good performance compared to the established baseline.

Table 1: Evaluation metrics of the best models for core, nowcasting, and transfer learning challenges, on the validation dataset. The values in parenthesis associated with the CSI scores indicate the thresholds used in evaluating the CSI scores.

| Metric | Core | Nowcasting | Transfer Learning |
|---|---|---|---|
| CSI (0.2) | 0.19466977 | 0.2181365 | 0.22192283 |
| CSI (1) | 0.061707865 | 0.0659472 | 0.0607376 |
| CSI (5) | 0.00639612 | 0.006552998 | 0.005987511 |
| CSI (10) | 0.00269436 | 0.001771289 | 0.001894928 |
| CSI (15) | 0.001170633 | 0.001328052 | 0.001374801 |
| CSI (overall) | 0.05332775 | 0.05874721 | 0.058383536 |
| HSS (overall) | 0.063846074 | 0.0722378 | 0.07194933 |

Table 2: Evaluation scores (CSI) of the best models for core, nowcasting, and transfer learning on the test dataset, in comparison with the leaderboard official baseline scores.

| CSI score | Core | Nowcasting | Transfer Learning |
|---|---|---|---|
| Leaderboard official baseline | 0.04443977 | 0.04821894 | 0.04585502 |
| PAUNet | 0.04877938 | 0.05689464 | 0.05754522 |

To quantitatively evaluate the downscaling and forecasting capabilities of PAUNet for precipitation, we employed a 2D spatial Fourier transform analysis. This involved computing the power spectral density (PSD) of the predicted precipitation maps and subsequently comparing them with the PSD of the actual target precipitation. First, the PSD is calculated for each of the predicted precipitation maps at respective time instances and averaged over all samples to get the mean PSD. Then, the mean PSD is radially averaged from the center and plotted versus the wavelength, as shown in Figure 3. The observed PSD comparison reveals distinctive characteristics, indicating that the model predictions exhibit higher power at larger wavelengths and reduced power at smaller wavelengths in contrast to the targets. This suggests that model produces much rain and substantially underestimates the extremes. This discrepancy suggests areas for improvement in the PAUNet model. Notably, a model with enhanced spatial information would demonstrate less deviation in the PSD spectrum compared to observations, underscoring the significance of PSD as a pertinent metric for comprehensive model evaluation in precipitation downscaling and forecasting.

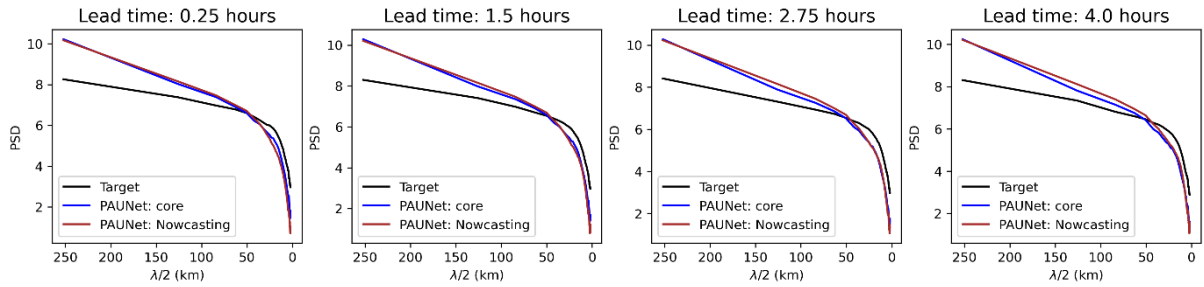

Fig. 3: Radially averaged mean PSD of precipitation, plotted against wavelength, for lead times of 0.25 hours, 1.5 hours, 2.75 hours, and 4 hours across all the samples of validation data. The figure contrasts the PSD profiles derived from target data and PAUNet model predictions, focusing on the validation dataset.

To provide a qualitative evaluation of the spatial representation captured by the model predictions in precipitation forecasting, Figure 4 illustrates spatial precipitation forecasts at four distinct lead times, comparing the outputs of the best models with the actual targets. Notably, observations reveal that both core and nowcasting models successfully replicate observed precipitation structures, albeit with an expanded spatial extent and diminished intensity. It is discernible, however, that the model predictions are slightly deviating from the fidelity observed in the targets. This discrepancy is particularly noteworthy as it may contribute to the higher power observed at larger wavelengths in the PSD analysis. Interestingly, despite these challenges, the nowcasting model exhibits a closer resemblance to the targets than the core model, showcasing a promising aspect in its spatial representation capabilities. These findings underscore the complexity of capturing high-resolution spatial patterns in precipitation forecasts and highlight areas for refinement in model performance.

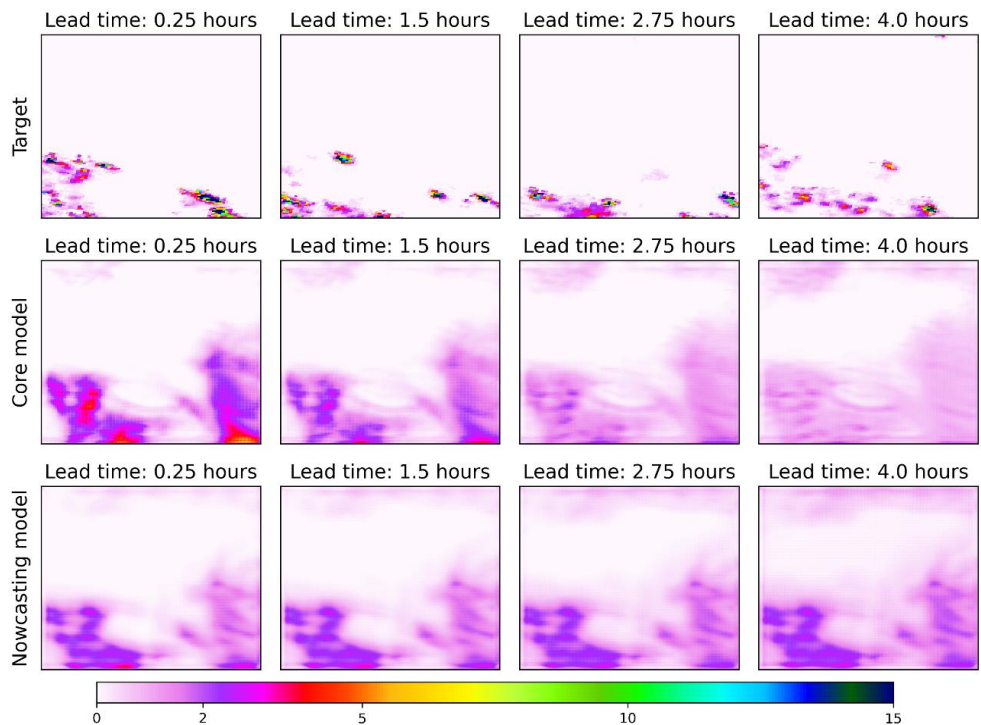

Fig. 4: An example illustration of precipitation forecast at four lead times from the best core and nowcasting models, compared to the targets.

# 5 Conclusions

In conclusion, our study introduced the Precipitation Attention-based U-Net (PAUNet) for high-resolution precipitation forecasting using the low-resolution satellite radiance data. Employed as a solution for the Weather4cast challenge, our proposed model exhibited superior performance, surpassing the official leaderboard baseline scores by notable margins of 10% in the core, 18% in nowcasting, and 25% in transfer learning challenges. A key contributing factor to the model's enhanced precipitation forecasting capability lies in the implementation of the exponential Focal Precipitation Loss (e-FPL) as the chosen loss function during training. Despite achieving improved results, it is acknowledged that the current model architecture has yet to undergo extensive experimentation, primarily constrained by time limitations. In future, we plan to explore improving the potential of PAUNet with diverse architectures, such as Generative Adversarial Networks (GANs) [14, 23], and investigate the value of providing orography information as an additional input [11]. The outcomes of our study not only underscore the efficacy of PAUNet in advancing precipitation forecasting but also point towards promising directions for continued exploration and enhancement in the evolving landscape of deep learning-based weather prediction methodologies.

## References

1. Rahmstorf, S., Coumou, D.: Increase of extreme events in a warming world. Proceedings of the National Academy of Sciences. 108, 17905–17909 (2011). https://doi.org/10.1073/PNAS.1101766108

2. Xu, L., Chen, N., Chen, Z., Zhang, C., Yu, H.: Spatiotemporal forecasting in earth system science: Methods, uncertainties, predictability and future directions. Earth-Science Reviews. 222, 103828 (2021). https://doi.org/10.1016/j.earscirev.2021.103828

3. Li, W., Gao, X., Hao, Z., Sun, R.: Using deep learning for precipitation forecasting based on spatio-temporal information: a case study. Clim Dyn. 58, 443–457 (2022). https://doi.org/10.1007/S00382-021-05916-4/TABLES/13

4. Lam, R., Sanchez-Gonzalez, A., Wilson, M., Wirnsberger, P., Fortunato, M., Alet, F., Ravuri, S., Ewalds, T., Eaton-Rosen, Z., Hu, W., Merose, A., Hoyer, S., Holland, G., Vinyals, O., Stott, J., Pritzel, A., Mohamed, S., Battagila, P.: Learning skillful medium-range global weather forecasting. Science. 0, eadi2336. https://www.science.org/doi/10.1126/science.adi2336

5. Pathak, J., Subramanian, S., Harrington, P., Raja, S., Chattopadhyay, A., Mardani, M., Kurth, T., Hall, D., Li, Z., Azizzadenesheli, K., Hassanzadeh, P., Kashinath, K., Anandkumar, A.: FourCastNet: A Global Data-driven High-resolution Weather Model using Adaptive Fourier Neural Operators. arXiv preprint arXiv:2202.11214 (2022). https://arxiv.org/pdf/2202.11214.pdf

6. Nguyen, T., Brandstetter, J., Kapoor, A., Gupta, J.K., Grover, A.: ClimaX: A foundation model for weather and climate. Proc Mach Learn Res. 202, 25904–25938 (2023)

7. Sønderby, C.K., Espeholt, L., Heek, J., Dehghani, M., Oliver, A., Salimans, T., Agrawal, S., Hickey, J., Kalchbrenner, N.: MetNet: A Neural Weather Model for Precipitation Forecasting. (2020)

8. Espeholt, L., Agrawal, S., Sønderby, C., Kumar, M., Heek, J., Bromberg, C., Gazen, C., Carver, R., Andrychowicz, M., Hickey, J., Bell, A., Kalchbrenner, N.: Deep learning for twelve hour precipitation forecasts. Nat Commun. 13, (2022). https://doi.org/10.1038/S41467-022-32483-X

9. Andrychowicz, M., Espeholt, L., Li, D., Merchant, S., Merose, A., Zyda, F., Agrawal, S., Kalchbrenner, N., Deepmind, G., Research, G.: Deep Learning for Day Forecasts from Sparse Observations. (2023)

10. Shi, X., Gao, Z., Lausen, L., Wang, H., Yeung, D.-Y., Wong, W.-K., Woo, W.-C., Kong Observatory, H.: Deep Learning for Precipitation Nowcasting: A Benchmark and A New Model. Adv Neural Inf Process Syst. 30, (2017)

11. Reddy, P.J., Matear, R., Taylor, J., Thatcher, M., Grose, M.: A precipitation downscaling method using a super-resolution deconvolution neural network with step orography. Environmental Data Science. 2, e17 (2023). https://doi.org/10.1017/EDS.2023.18

12. Shi, X., Chen, Z., Wang, H., Yeung, D.-Y., Wong, W.-K., Woo, W.-C., Kong Observatory, H.: Convolutional LSTM Network: A Machine Learning Approach for Precipitation Nowcasting.

13. Taylor, J.A., Larraondo, P., de Supinski, B.R.: Data-driven global weather predictions at high resolutions. The International Journal of High Performance Computing Applications. 36(2), 130-140 (2022). https://doi.org/10.1177/10943420211039818

14. Ravuri, S., Lenc, K., Willson, M., Kangin, D., Lam, R., Mirowski, P., Fitzsimons, M., Athanassiadou, M., Kashem, S., Madge, S., Prudden, R., Mandhane, A., Clark, A., Brock, A., Simonyan, K., Hadsell, R., Robinson, N., Clancy, E., Arribas, A., Mohamed, S.: Skilful precipitation nowcasting using deep generative models of radar. Nature. 597, 672–677 (2021). https://doi.org/10.1038/S41586-021-03854-Z

15. Gruca, A., Serva, F., Lliso, L., Rípodas, P., Calbet, X., Herruzo, P., Pihrt, J., Raevskyi, R., Šimánek, P., Choma, M., Li, Y., Dong, H., Belousov, Y., Polezhaev, S., Pulfer, B., Seo, M., Kim, D., Shin, S., Kim, E., Ahn, S., Choi, Y., Park, J., Son, M., Cho, S., Lee, I., Kim, C., Kim, T., Kang, S., Shin, H., Yoon, D., Eom, S., Shin, K., Yun, S.-Y., Le Saux, B., Kopp, M.K., Hochreiter, S., Kreil, D.P., Ciccone, M., Stolovitzky, G., Albrecht, J., sro, M.: Weather4cast at NeurIPS 2022: Super-Resolution Rain Movie Prediction under Spatio-temporal Shifts. In: Proceedings of Machine Learning Research. pp. 292–312 (2023)

16. Herruzo, P., Gruca, A., Lliso, L., Calbet, X., Ripodas, P., Hochreiter, S., Kopp, M., Kreil, D.P.: High-resolution multi-channel weather forecasting - First insights on transfer learning from the Weather4cast Competitions 2021. Proceedings - 2021 IEEE International Conference on Big Data, Big Data 2021. 5750–5757 (2021). https://doi.org/10.1109/BIGDATA52589.2021.9672063

17. Gruca, A., Herruzo, P., Rípodas, P., Kucik, A., Briese, C., Kopp, M.K., Hochreiter, S., Ghamisi, P., Kreil, D.P.: CDCEO'21 - First Workshop on Complex Data Challenges in Earth Observation. International Conference on Information and Knowledge Management, Proceedings. 4878–4879 (2021). https://doi.org/10.1145/3459637.3482044

18. Ronneberger, O., Fischer, P., Brox, T.: U-net: Convolutional networks for biomedical image segmentation. In: Medical Image Computing and Computer-Assisted Intervention–MICCAI 2015: 18th International Conference, Munich, Germany, October 5-9, 2015, Proceedings, Part III 18. pp. 234–241. Springer (2015)

19. He, K., Zhang, X., Ren, S., Sun, J.: Deep residual learning for image recognition. In: Proceedings of the IEEE conference on computer vision and pattern recognition. pp. 770–778 (2016)

20. Vaswani, A., Shazeer, N., Parmar, N., Uszkoreit, J., Jones, L., Gomez, A.N., Kaiser, Ł., Polosukhin, I.: Attention is all you need. Adv Neural Inf Process Syst. 30, (2017)

21. Li, Y., Dong, H., Fang, Z., Weyn, J., Luferenko, P.: Super-resolution Probabilistic Rain Prediction from Satellite Data Using 3D U-Nets and EarthFormers. arXiv preprint arXiv:2212.02998. (2022)

22. Park, J., Son, M., Cho, S., Lee, I., Kim, C.: RainUNet for Super-Resolution Rain Movie Prediction under Spatio-temporal Shifts. arXiv preprint arXiv:2212.04005. (2022)

23. Huang, P.-C., Chen, Y.-L., Liou, Y.-S., Tsai, B.-C., Wu, C.-C., Hsu, W.H.: STAMINA (Spatial-Temporal Aligned Meteorological INformation Attention) and FPL (Focal Precip Loss): Advancements in Precipitation Nowcasting for Heavy Rainfall Events. In: Proceedings of the 32nd ACM International Conference on Information and Knowledge Management. pp. 854–863. Association for Computing Machinery, New York, NY, USA (2023)

24. Abadi, M., Agarwal, A., Barham, P., Brevdo, E., Chen, Z., Citro, C., Corrado, G.S., Davis, A., Dean, J., Devin, M., Ghemawat, S., Goodfellow, I., Harp, A., Irving, G., Isard, M., Jia, Y., Jozefowicz, R., Kaiser, L., Kudlur, M., Levenberg, J., Mané, D., Monga, R., Moore, S., Murray, D., Olah, C., Schuster, M., Shlens, J., Steiner, B., Sutskever, I., Talwar, K., Tucker, P., Vanhoucke, V., Vasudevan, V., Viégas, F., Vinyals, O.,

Warden, P., Wattenberg, M., Wicke, M., Yu, Y., Zheng, X., Research, G.: TensorFlow: Large-Scale Machine Learning on Heterogeneous Distributed Systems, (2015)

25. Sergeev, A., Del Balso, M.: Horovod: fast and easy distributed deep learning in TensorFlow. (2018). https://doi.org/10.48550/arxiv.1802.05799